\documentclass[aps,prl,reprint,groupedaddress]{revtex4-1}
\usepackage{gensymb}
\usepackage{graphicx}
\begin{document}

\title{Nematic Superconductivity in Cu$_{1.5}$(PbSe)$_{5}$(Bi$_{2}$Se$_{3}$)$_{6}$}


\author{Lionel Andersen}
\author{Zhiwei Wang}
\author{Thomas Lorenz}
\author{Yoichi Ando}
\email[]{ando@ph2.uni-koeln.de}
\affiliation{Physics Institute II, University of Cologne, 50937 K\"oln, Germany}

\date{\today}

\begin{abstract}
After the discovery of nematic topological superconductivity in Cu$_{x}$Bi$_{2}$Se$_{3}$, carrier-doped topological insulators are established as a fertile ground for topological superconductors. The superconductor Cu$_{1.5}$(PbSe)$_{5}$(Bi$_{2}$Se$_{3}$)$_{6}$ (CPSBS) contains Bi$_{2}$Se$_{3}$ blocks as a constitutional unit, but its superconducting gap appears to have nodes [S. Sasaki {\it et al.}, Phys. Rev. B {\bf 90}, 220504 (2014)], which is in contrast to the fully-opened gap in Cu$_{x}$Bi$_{2}$Se$_{3}$ and the relation between the two superconductors remained an open question. Here we report our observation of clear two-fold symmetry in the in-plane magnetic-field-direction dependencies of the upper critical field and of the specific heat of CPSBS, where the direction of the maxima, which is different from that in Cu$_{x}$Bi$_{2}$Se$_{3}$, indicates that the gap nodes are located in the mirror plane of the crystal lattice. This means that the topological nematic state with mirror-symmetry-protected nodes is realized in CPSBS. 
\end{abstract}


\pacs{74.25.Bt, 74.25.Dw, 74.20.Mn, 03.65.Vf}




\maketitle



The search for concrete materials to realize novel topological states of matter is an exciting frontier in condensed matter physics \cite{Hasan2010, Qi2011, Ando2013}. In that search, topological superconductors attract particular attention due to the potential appearance of exotic quasiparticles called Majorana fermions at their boundaries \cite{Qi2011, Alicea2012, Beenakker2013, Elliott2015, Sato2017}. The superconductors derived from topological insulators (TIs) are expected to be a fertile ground in this respect, owing to the strong spin-orbit coupling which may give rise to an unconventional momentum-dependent superconducting gap even for the isotropic pairing force coming from conventional electron-phonon interactions \cite{Fu2010, Ando2015}.

The first of such materials was Cu$_x$Bi$_2$Se$_3$ \cite{Hor2010}, which is synthesized by intercalating Cu into the van der Waals gap of the prototypical TI material Bi$_2$Se$_3$. Cu$_x$Bi$_2$Se$_3$ shows superconductivity with $T_c \simeq$ 3 K for $x \simeq$ 0.3, and early point-contact spectroscopy measurements pointed to the occurrence of topological superconductivity associated with surface Majorana fermions \cite{Sasaki2011}. Recent measurements of its bulk superconducting properties have elucidated \cite{Matano2016,Yonezawa2017} that it realizes a topological superconducting state which spontaneously breaks in-plane rotational symmetry in a two-fold symmetric manner, even though the crystal lattice symmetry is three-fold. Such an unconventional state is consistent with one of the four possible superconducting states constrained by the $D_{3d}$ lattice symmetry of Bi$_2$Se$_3$ \cite{Fu2010, Ando2015}; this state, named $\Delta_{4x}$ or $\Delta_{4y}$ state depending on the direction of nodes or gap minima, is characterized by a nematic order parameter and hence is called a {\it nematic superconducting state} \cite{Fu2014}. It was reported that Sr$_x$Bi$_2$Se$_3$ \cite{Liu2015} and Nb$_x$Bi$_2$Se$_3$ \cite{Qiu2015} also realize the nematic superconducting state \cite{Nikitin2016, Pan2016, Shen2017, Asaba2017, Du2017, Kuntsevich2018, Smylie2018}.

An interesting superconducting compound related to Cu$_x$Bi$_2$Se$_3$ is Cu$_{x}$(PbSe)$_{5}$(Bi$_{2}$Se$_{3}$)$_{6}$ (CPSBS), which was discovered in 2014 \cite{Sasaki2014}. Its parent compound (PbSe)$_{5}$(Bi$_{2}$Se$_{3}$)$_{6}$ (PSBS) can be viewed as a natural heterostructure formed by a stack of two-quintuple-layer (QL) Bi$_{2}$Se$_{3}$ units alternating with one-bilayer PbSe units \cite{Nakayama2012, Fang2014, Segawa2015, Nakayama2015, Momida2018}. Since the binary compound PbSe is a topologically trivial insulator, PSBS consists of ultra-thin TI layers separated by trivial-insulator layers. When Cu is intercalated into the van der Waals gap in the Bi$_{2}$Se$_{3}$ unit of PSBS, superconductivity with $T_c$ = 2.8~K shows up and nearly 100\% superconducting volume fraction can be obtained for $x \simeq$ 1.5. Since the structural unit responsible for superconductivity in CPSBS is essentially Cu$_x$Bi$_2$Se$_3$, one would expect the same unconventional superconducting state to be realized in CPSBS. Nevertheless, there is a marked difference between the two compounds: whereas there is strong evidence that CPSBS has gap nodes \cite{Sasaki2014}, Cu$_x$Bi$_2$Se$_3$ is fully gapped \cite{Kriener2011}. Hence, it is important to clarify the nature of the nodal superconducting state in CPSBS. 

In this paper, we report our discovery of two-fold symmetry in the upper critical field $H_{c2}$ and the specific heat $c_p$ in their dependencies on the magnetic-field direction in the basal plane. The pattern of the two-fold symmetry indicates that the gap nodes are lying in the mirror plane of the crystal, suggesting that the $\Delta_{4x}$ state with symmetry-protected nodes is realized in CPSBS. This is in contrast to the $\Delta_{4y}$ state realized in Cu$_x$Bi$_2$Se$_3$, in which the nodes are not protected by symmetry and thus are lifted to form gap minima. We discuss that the likely cause of the $\Delta_{4x}$ state is the weak distortion of the Bi$_2$Se$_3$ lattice imposed by the PbSe units. This establishes CPSBS as a nematic topological superconductor with symmetry-protected nodes.

High-quality PSBS single crystals were grown by using a modified Bridgman method following Refs. \cite{Sasaki2014, Segawa2015}. X-ray Laue images were used for identifying the crystallographic $a$ axis upon cutting the pristine crystals, which were then electrochemically treated to intercalate Cu following the recipe of Kriener {\it et al.} \cite{Kriener2011b}, and the superconductivity was activated by annealing. The precise $x$ values determined by the weight change \cite{Kriener2011b} was 1.47 for the two samples presented here. The superconducting shielding fraction (SF) of the samples was measured in a commercial SQUID magnetometer. Further experimental details are given in the Supplemental Material \cite{Suppl}.

\begin{figure}[t]
	\centering
	\includegraphics[width=1.0\linewidth]{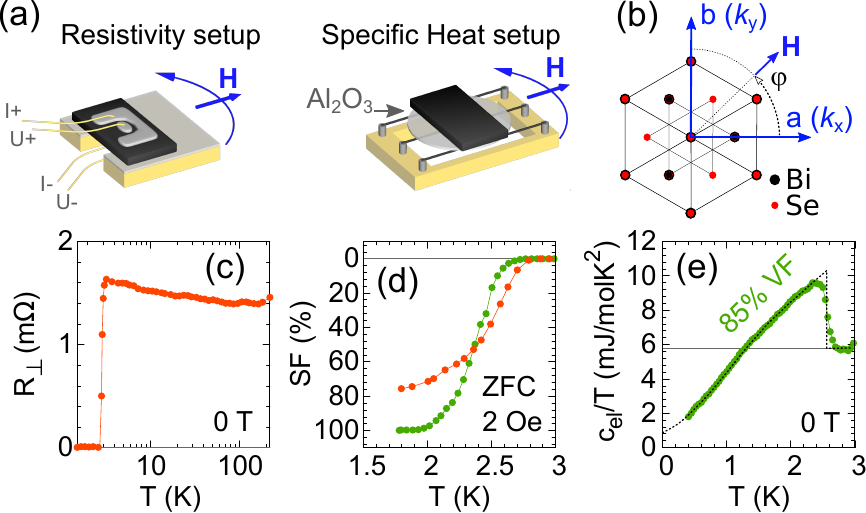}
	\caption{(a) Schematic pictures of the setups to measure the in-plane magnetic-field-direction dependencies of the out-of-plane resistance $R_{\perp}$ and the specific heat $c_p$. (b) The monoclinic $a$ axis of CPSBS lies in the mirror plane of the Bi$_2$Se$_3$ layers which nearly preserve the trigonal symmetry. (c) Temperature dependence of $R_{\perp}$ in sample A used for the resistive $H_{c2}$ measurements. (d) The zero-field-cooled (ZFC) magnetization data showing shielding fractions of 75~\% and 104~\% in samples A (red) and B (green), respectively.  (e) Temperature dependence of the electronic specific heat $c_{\rm el}$ in 0 T for sample B used for detailed $c_{\rm el}(H)$ measurements; the solid line is the theoretical curve for a line-nodal superconducting gap \cite{Won1994} assuming a superconducting volume fraction of 85~\%. To facilitate the comparison with Cu$_x$Bi$_2$Se$_3$, the molar volume is taken here for 1 mole of Bi$_2$Se$_3$.}
\end{figure}

To elucidate the possible in-plane anisotropy of the superconducting state, we employed the measurements of both the out-of-plane resistance $R_{\perp}$ and the specific heat $c_p$ in various orientations of the in-plane magnetic field $H$ [see Fig. 1(a) for configurations]. From the $R_{\perp}(H)$ data, the upper critical field $H_{c2}$ was extracted by registering the field where 50\% of the normal-state resistance is recovered. Note that $R_{\perp}$ measurements do not impose any in-plane anisotropy associated with the current direction. 
The $c_p$ measurements were performed with a standard relaxation method using a home-built calorimeter \cite{Suppl} optimized for small heat capacities.
Both measurements were done in a split-coil magnet with a $^3$He insert (Oxford Instruments Heliox), with which the magnetic-field direction with respect to the sample holder can be changed with a high accuracy ($\pm1\degree$) by rotating the insert in the magnet. With a manual second rotation axis on the cold finger, measurements with $H$ rotating in either the $ab$ or $ac^*$ plane were possible (note that $\vec{c^*} \parallel \vec{a} \times \vec{b}$ \cite{Segawa2015, Suppl}). 
We estimate the possible misalignment of the magnetic field to be $\pm 2\degree$. The two samples used for measuring $R_{\perp}$ and $c_p$ shown here presented the SF of 75\% and 104\%, respectively [Fig. 1(d)]. Here, no demagnetization correction is applied, since the magnetic field was applied parallel to the wide face of the platelet-shaped samples so that the demagnetization factor was $>$ 0.95.

\begin{figure}[t]
	\centering
	\includegraphics[width=1.0\linewidth]{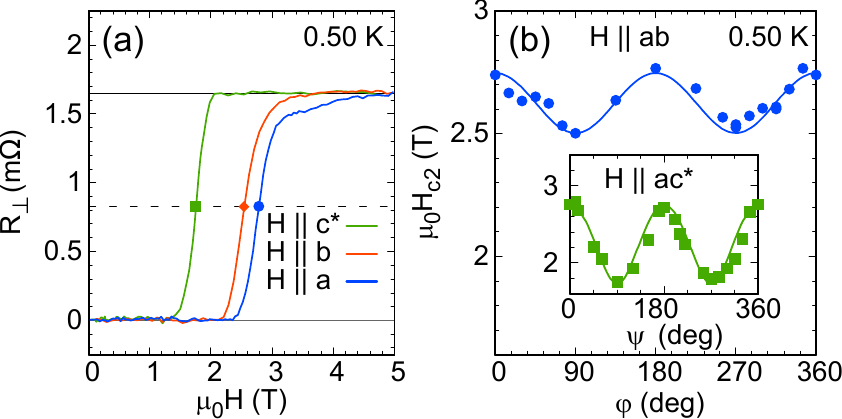}
	\caption{(a) $R_{\perp}(H)$ curves measured for three principal directions of applied magnetic fields, showing a clear difference in $H_{c2}$. (b) Magnetic-field-direction dependencies of $H_{c2}$ obtained from the $R_{\perp}(H)$ data in the in-plane rotation (main panel) and the out-of-plane rotation (inset); the angles $\varphi$ and $\psi$ are measured from the $a$ axis.} 
\end{figure}


The temperature dependence of $R_{\perp}$ presents a weak upturn below $\sim$100 K [Fig. 1(c)], which reflects the quasi-two-dimensional (2D) electronic states of CPSBS. The $R_{\perp}(H)$ curves measured at 0.5 K with the applied magnetic field in the three orthogonal directions, $a$, $b$, and $c^*$ axes, are shown in Fig. 2(a). One can immediately see that $H_{c2}$ for the three magnetic-field directions are different; the smallest value for $H \parallel c^*$ is a consequence of the quasi-2D nature and was already reported \cite{Sasaki2014}, but the anisotropy between  $H \parallel a$ and $H \parallel b$ is a new observation. The precise in-plane magnetic-field-direction dependence of $H_{c2}$ at 0.5 K is shown in the main panel of Fig. 2(b), where one can see clear two-fold symmetry with maxima at $H \parallel a$ and the variation $\Delta H_{c2}^{\parallel}$ of $\sim$0.25 T. As explained in detail in the Supplemental Material \cite{Suppl}, the $a$ axis in CPSBS is parallel to the mirror plane and hence the direction of $H_{c2}$ maxima is 90$^{\circ}$ rotated from that in Cu$_x$Bi$_2$Se$_3$ \cite{Yonezawa2017}. We note that anisotropic $H_{c2}$ measurements with the current along the $b$ axis were also performed, and $H_{c2}$ was not affected by the current direction \cite{Suppl}. Also, the $H_{c2}$ anisotropy in $R_{\perp}(H)$ was reproduced in one more sample \cite{Suppl}.

For comparison, the magnetic-field-direction dependence of $H_{c2}$ at 0.5 K in the $ac^*$ plane is shown in the inset of Fig. 2(b), where the magnitude of the variation in $H_{c2}$, $\Delta H_{c2}^{\perp}$, is about 1.0 T. This $\Delta H_{c2}^{\perp}$ value means that, for the observed two-fold in-plane anisotropy with $\Delta H_{c2}^{\parallel} \sim$ 0.25 T to be ascribed to an accidental $c^*$-axis component of $H$, a sample misalignment of $\sim$30$\degree$ would be necessary. This is obviously beyond the possible error in our experimental setup, and one can conclude that the two-fold in-plane anisotropy is intrinsic.


Due to the volume sensitivity, the $c_p(T)$ data provides a better estimate of the superconducting volume fraction (VF) than the diamagnetic SF. After subtracting the phononic contribution \cite{Suppl}, the electronic specific heat $c_{\rm el}$ shows a clear anomaly associated with the superconducting transition; Fig. 1(e) shows a plot of $c_{\rm el}/T$ vs $T$, which is fitted with a line-nodal gap theory \cite{Won1994} used for CPSBS in Ref. \cite{Sasaki2014}. 
This fitting yields the superconducting VF of 85\% for this sample, which is used for further $c_{\rm el}$ measurements.

\begin{figure}[t] 
	\centering
	\includegraphics[width=1.0\linewidth]{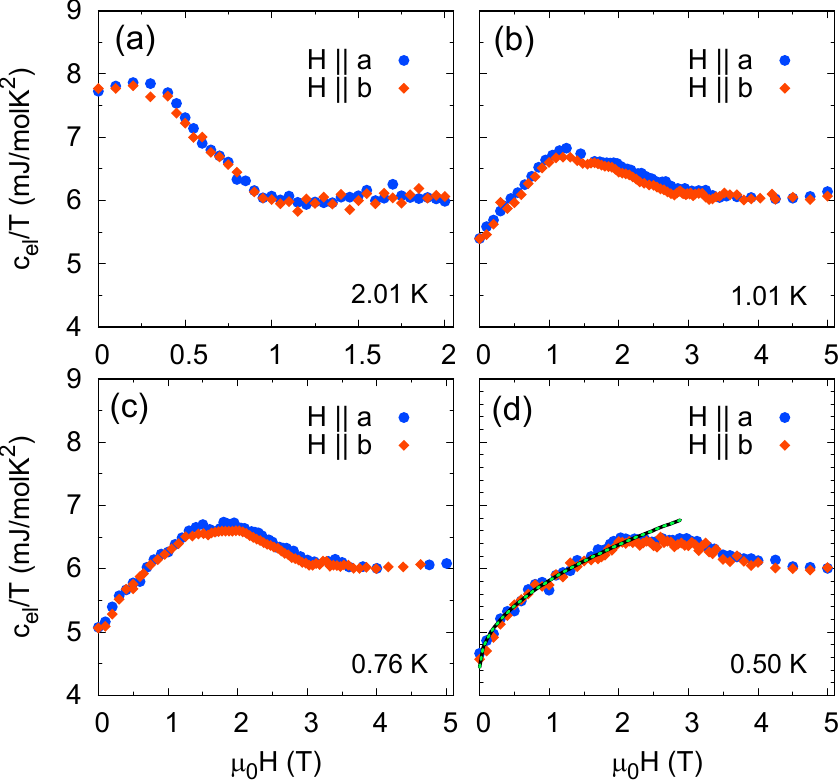}
	\caption{Magnetic-field dependencies of $c_{\rm el}/T$ at (a) 2.01 K, (b) 1.01 K, (c) 0.76 K, and (d) 0.50 K measured in $H \parallel a$ and $H \parallel b$. The dashed line in panel (d) shows the $\sqrt{H}$ behavior expected for a superconducting gap with line nodes, as was already reported in Ref. \cite{Sasaki2014}.}
\end{figure}

\begin{figure}[t]
\centering
\includegraphics[width=1.0\linewidth]{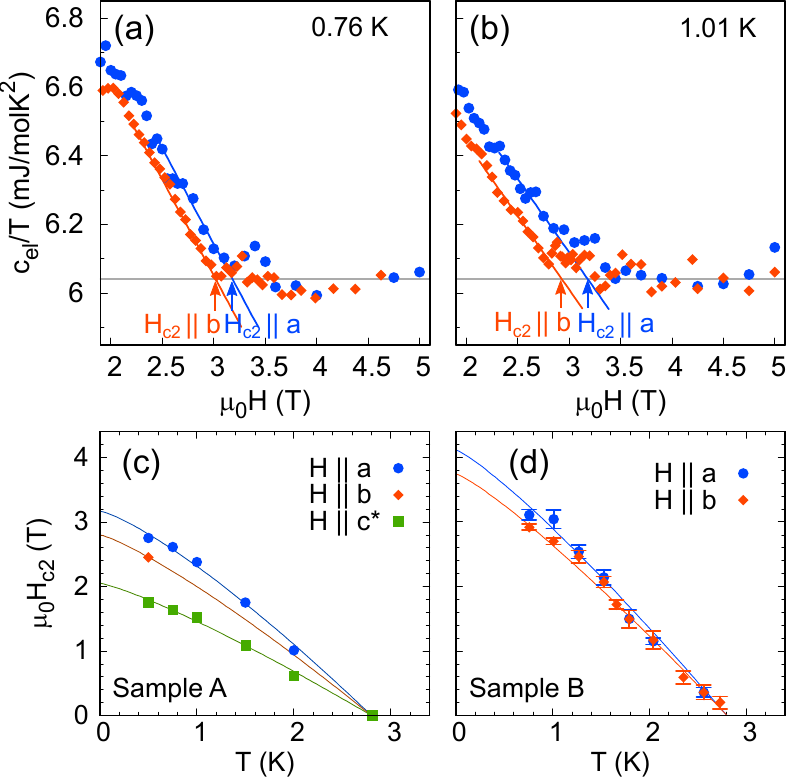}
\caption{(a),(b) Magnification of the $c_{\rm el}(H)$ behavior near the $H_{c2}$ at (a) 0.76 K and (b) 1.01 K for $H \parallel a$ and $H \parallel b$, showing how the $H_{c2}$ values were extracted. (c),(d) Temperature dependencies of $H_{c2}$ extracted from (c) the middle-point in the $R_{\perp}(H)$ transitions in sample A and (d) disappearance of the superconducting contribution in $c_{\rm el}$ in sample B, for the principal magnetic-field directions; the solid lines are fits to the empirical $\sim 1-(T/T_c)^{1.2}$ dependence.}
\end{figure}

The magnetic-field dependencies of $c_{\rm el}$ at various temperatures for both $H \parallel a$ and $H \parallel b$ are shown in Fig. 3; the data presented here are after subtracting the Schottky anomaly \cite{Suppl} by using the same $g$-factor as that reported in Ref. \cite{Sasaki2014}. 
One can see that at 2.01 and 1.01 K, $c_{\rm el}$ changes little above a certain $H$ value, which we identify as $H_{c2}$. 
However, at lower temperature ($\lesssim$ 0.5 K) the change in the $c_{\rm el}(H)/T$ behavior across $H_{c2}$ becomes less evident and we lose the sensitivity to determine $H_{c2}$. As a result, the in-plane anisotropy in $H_{c2}$ is best visible in $c_{\rm el}$ at intermediate temperatures around 1 K [Figs. 4(a) and 4(b)]. 
In our analysis of $c_{\rm el}(H)$, $H_{c2}$ was determined as the crossing point of the two linear fittings of the $c_{\rm el}/T$ vs $H$ data below and above $H_{c2}$ as shown in Figs. 4(a) and 4(b); here, one can see that the difference in $H_{c2}$ for $H \parallel a$ and $H \parallel b$ is better discernible at 1.01 K with $\Delta H_{c2} \sim$ 0.34 T than at 0.76 K. Importantly, $H_{c2}$ is larger for $H \parallel a$, which is consistent with the result of the $R_{\perp}(H)$ measurements. 

The temperature dependencies of $H_{c2}$ extracted from $R_{\perp}(H)$ and $c_{\rm el}(H)$ for the principal magnetic-field orientations are plotted in Figs. 4(c) and 4(d), respectively. The absolute values of $H_{c2}$ in the two panels are different, mainly because Fig. 4(c) shows the mid-point of the transition while Fig. 4(d) shows the complete suppression. Nevertheless, the in-plane anisotropy is consistently found in the $R_{\perp}(H)$ and $c_{\rm el}(H)$ results. 
In Figs. 4(c) and 4(d), the $H_{c2}(T)$ data are fitted empirically with $H_{c2}(T) = H_{c2}(0) \left[1-(T/T_c)^{a} \right]$ with $a \approx$ 1.2; the inapplicability of the conventional Werthamer-Helfand-Hohenberg theory for $H_{c2}(T)$ was already reported for CPSBS and was discussed to be a possible consequence of unconventional pairing \cite{Sasaki2014}.

\begin{figure}[t]
	\centering
	\includegraphics[width=1.0\linewidth]{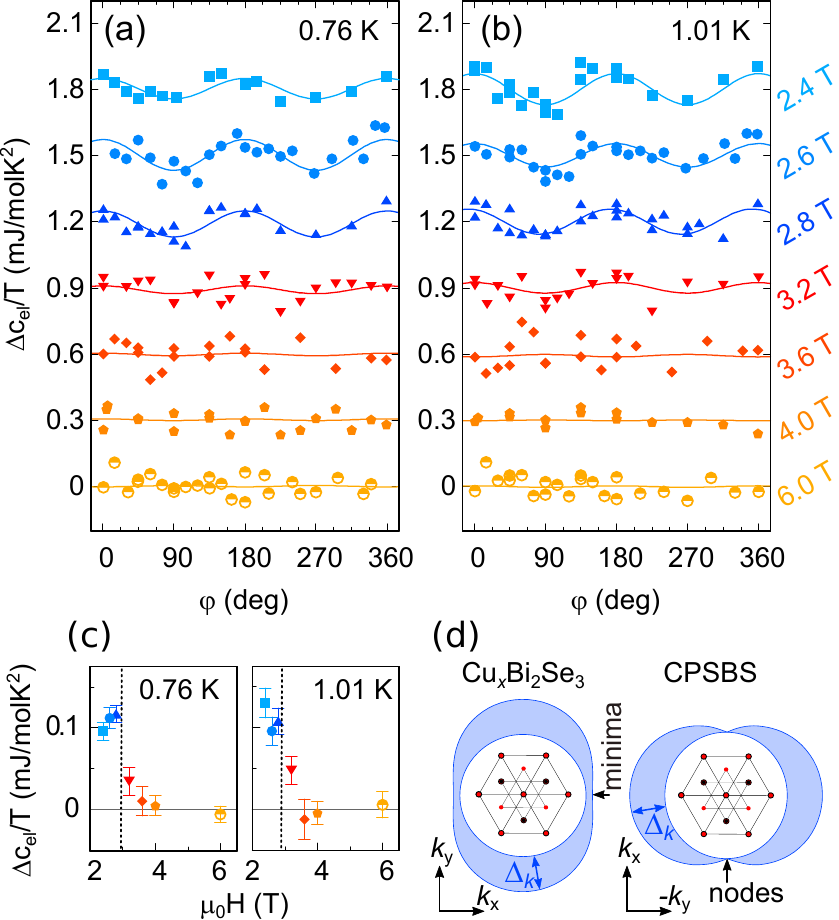}
	\caption{(a),(b) Change in $c_{\rm el}/T$ as a function of the angle $\varphi$ of the applied in-plane magnetic field at constant strengths of $H$ across $H_{c2}$ (2.4 -- 6.0 T, the data are shifted for clarity) at 0.76 and 1.01 K, presenting two-fold symmetric oscillations at $H < H_{c2}$.
(c) Dependence of the oscillation amplitude on the strength of $H$ at 0.76 and 1.01 K, demonstrating its quick disappearance above $H_{c2}$.
(d) Schematic pictures of the $\Delta_{4y}$ and $\Delta_{4x}$ gaps, which are realized in Cu$_x$Bi$_2$Se$_3$ and CPSBS, respectively, in relation to the Bi$_2$Se$_3$ lattice.}
\end{figure}

To supplement the conclusion from $H_{c2}$, we have also measured the detailed magnetic-field-direction dependence of $c_{\rm el}$ at 0.76 and 1.01 K in various strengths of the in-plane magnetic field from 2.4 to 6.0 T [Figs. 5(a) and 5(b)]. One can clearly see two-fold symmetric variations where the maxima occur at $H \parallel a$ for $H < H_{c2}$, but the anisotropy quickly disappears for $H > H_{c2}$ [see also Fig. 5(c)]. This disappearance of the anisotropy in the normal state strongly supports the interpretation that the anisotropy is due to the nematicity which arises spontaneously in the superconducting state. It also demonstrates that the observed $c_p$ anisotropy cannot be due to some $g$-factor anisotropy which might show up through the Schottky anomaly.

It is useful to note that in Cu$_x$Bi$_2$Se$_3$, a sign change in the magnetic-field-direction dependence of $c_{\rm el}$ was observed \cite{Yonezawa2017}; namely, the maxima in $c_{\rm el}$ were observed for $H$ normal to a mirror plane  at high $T$ and/or high $H$, but at low $T$ and low $H$, $c_{\rm el}$ presented minima in this direction. Such a switching behavior was explained as a result of the competition between the Doppler-shift effect and the vortex-scattering effect discussed by Vorontsov and Vekhter (VV) \cite{Vorontsov2006}. According to VV, the latter effect is dominant at higher $H$ at any temperature in a nodal superconductor and causes the maxima in $c_{\rm el}$ to appear for $H$ in the nodal direction. 

In view of the VV theory, the two-fold in-plane anisotropy in CPSBS with maxima in $c_{\rm el}$ appearing for $H \parallel a$ near $H_{c2}$ points to the realization of the $\Delta_{4x}$-type superconducting gap, which has gap nodes in the mirror plane [see Fig. 5(d)]. This conclusion is different from that for Cu$_x$Bi$_2$Se$_3$ \cite{Yonezawa2017}, where the $\Delta_{4y}$ state is realized. Note that the direction of maxima in $H_{c2}(\varphi)$ \cite{Venderbos2016} is also consistent with the $\Delta_{4x}$ gap in CPSBS and with the $\Delta_{4y}$ gap in Cu$_x$Bi$_2$Se$_3$.
While the $\Delta_{4x}$ state was originally predicted for the three-dimensional ellipsoidal Fermi surface of Bi$_2$Se$_3$ to have point nodes \cite{Fu2010}, the quasi-2D nature of the Fermi surface in CPSBS \cite{Nakayama2015} extends the original point nodes into line nodes along the $c^*$ direction, at least in the extreme 2D limit \cite{Hashimoto2014}. 
As pointed out by Fu \cite{Fu2014}, the nodes in the $\Delta_{4x}$ state are protected by mirror symmetry, which explains why CPSBS is a nodal superconductor despite its essential similarity to Cu$_x$Bi$_2$Se$_3$.


It is useful to mention that the crystallographic symmetry of CPSBS belongs to the monoclinic space group $C2/m$ \cite{Fang2014, Segawa2015, Momida2018}, which means that the lattice is actually two-fold symmetric \cite{Foot2}. The monoclinic nature arises from the fact that PSBS is a heterostructure of two dissimilar crystal symmetries, the trigonal lattice of Bi$_2$Se$_3$ and the square lattice of PbSe (see \cite{Suppl} for details). The lowering of the symmetry makes one of the three equivalent mirror planes in Bi$_2$Se$_3$ to be the only mirror plane, which contains the monoclinic $a$ axis; in fact, there is a weak but finite uniaxial distortion \cite{Suppl} in the Bi$_2$Se$_3$ QL units in PSBS \cite{Fang2014, Segawa2015}.
According to the theory \cite{Fu2014, Venderbos2016}, under the constraint of the $D_{3d}$ point group, an odd-parity superconducting state which breaks in-plane rotation symmetry must obey $E_u$ symmetry and in general has a nematic gap function $\Delta(\mathbf{k}) = \eta_1 \Delta_{4x} + \eta_2 \Delta_{4y}$, where the two nodal gap functions $\Delta_{4x}$ and $\Delta_{4y}$ are degenerate and $\vec{\eta} = (\eta_1, \eta_2)$ can be viewed as the nematic director. This is why the $E_u$ state is called nematic. However, for the physical properties to present a two-fold anisotropy, a uniaxial symmetry-breaking perturbation is necessary \cite{Venderbos2016}. In CPSBS, the weak uniaxial lattice distortion, which leads to the $C2/m$ symmetry, is apparently responsible for lifting the degeneracy between $\Delta_{4x}$ and $\Delta_{4y}$ and make the nematic director to take the definite direction $\vec{\eta}$ = (1,0). Such a situation is rather similar to that realized in the high-$T_c$ cuprate YBa$_2$Cu$_3$O$_y$, in which a tiny orthorhombic distortion dictates the orientation of the spontaneously-formed nematic state \cite{Ando2002}, although the nematicity is about the normal state in YBa$_2$Cu$_3$O$_y$ while it is about the superconducting state in CPSBS.

We note that the ARPES measurements on superconducting CPSBS found no clear evidence for two-fold-symmetric Fermi surface distortion within the experimental error of $\sim$2 \% \cite{Nakayama2015}. Hence, the possible anisotropy in the Fermi velocity $v_{\rm F}$ cannot be large enough to directly account for the observed two-fold anisotropy in $H_{c2}$ of $\sim$10\%, but it must be responsible for lifting the degeneracy in the nematic state. The emerging picture is that the microscopic physics of electrons in doped Bi$_2$Se$_3$ chooses the $E_u$ superconducting state to be the most energetically favorable, and then a symmetry-breaking perturbation sets the direction of $\vec{\eta}$, so that a two-fold anisotropy shows up \cite{noteMR}. In this regard, there is a report that two-fold anisotropy in Sr$_x$Bi$_2$Se$_3$ correlates with a weak structural distortion \cite{Kuntsevich2018}. Interestingly, existence of gap nodes has been suggested for Nb$_x$Bi$_2$Se$_3$ \cite{Smylie2016, Smylie2017}, which implies that the symmetry-breaking perturbation in Nb$_x$Bi$_2$Se$_3$ is different from that in Cu$_x$Bi$_2$Se$_3$ and prompts $\vec{\eta}$ to take (1,0). 

In summary, we found that both the $H_{c2}$ and the $c_p$ of superconducting CPSBS present two-fold-symmetric in-plane anisotropy with maxima occurring for $H \parallel a$. This points to the realization of the $\Delta_{4x}$-type superconducting gap associated with symmetry-protected line nodes extending along the $c^*$ direction. Hence, CPSBS is a nematic topological superconductor differing from Cu$_x$Bi$_2$Se$_3$ in the orientation of the nematic director.

\begin{acknowledgments}
We thank T. Sato for re-analyzing the raw data of Ref. \cite{Nakayama2015} to check for possible Fermi-surface anisotropy in CPSBS. We also thank Y. Vinkler-Aviv for helpful discussions about the point group. 
This work was funded by the Deutsche Forschungsgemeinschaft (DFG, German Research Foundation) - Project number 277146847 - CRC 1238 (Subprojects A04 and B01).
\end{acknowledgments}


\begin{thebibliography}{41}
\bibitem [{Foo()}]{Foot2}%
  \BibitemOpen
  \href@noop {} {}\bibinfo {note} {The $C2/m$ symmetry is compatible with both
  the $\Delta_{4x}$ and $\Delta_{4y}$ gaps, which are derived from the $E_u$
  representation of the $D_{3d}$ point group.}\BibitemShut {Stop}%
\end{thebibliography}

%


\clearpage
\onecolumngrid

\renewcommand{\thefigure}{S\arabic{figure}} 
\renewcommand{\thetable}{S\arabic{table}}

\setcounter{figure}{0}

\begin{flushleft} 
{\Large {\bf Supplemental Material}}
\end{flushleft} 
\vspace{2mm}

\begin{flushleft} 
{\bf S1. Sample Synthesis}
\end{flushleft}

\begin{figure}[b]
	\centering
	\includegraphics[width=0.5\linewidth]{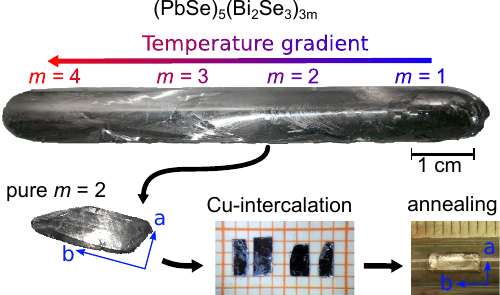}
	\caption{Photograph of a cm-sized sample rod containing the $m$ = 1 -- 4 phases of the (PbSe)$_{5}$(Bi$_{2}$Se$_{3}$)$_{3m}$ homologous series. The phases with different $m$ values are identified by XRD analyses. Single crystals of the $m$ = 2 phase with masses of 5 -- 70~mg were used for electrochemical Cu-intercalation and successive annealing to activate the superconductivity.}
\end{figure}

The parent compound  (PbSe)$_{5}$(Bi$_{2}$Se$_{3}$)$_{6}$ (PSBS) is a member of the homologous series  (PbSe)$_{5}$(Bi$_{2}$Se$_{3}$)$_{3m}$ with $m$ = 2 \cite{Segawa2015}. For this homologous series, compounds with $m$ = 1, 2, 3, and 4 have been synthesized \cite{Segawa2015}.
High-quality single crystals of PSBS were grown by using a combination of a modified Bridgman method and a self-flux method \cite{Segawa2015} because of the complex phase diagram \cite{Shelimova2010}. Essentially, we followed the recipe described in Refs.~\cite{Sasaki2014,Segawa2015}, but in order to increase the size of the single crystals of the $m$ = 2 phase, we have modified the starting composition of PbSe:Bi$_{2}$Se$_{3}$ from the ratio 45:55 to 38:62. In addition, we have reduced the cooling rate of the melt during the growth to 3 \degree C/day (710 \degree C to 640 \degree C in 560 h). Then the grown rod was fast cooled to room temperature. 

An optimum result was obtained with $\sim$7~g sample rods of $\sim$7~cm in length and $\sim$8~mm in diameter. A picture of a successfully grown rod, which contains phases from $m$ = 1 to 4, is shown in Fig. S1. Such a rod contains a small polycrystalline part at the tip (where the crystallization nucleates) and a quick crossover to single-crystalline sheet-like domains which span the whole width. The phases with different $m$ values are easily identified by X-ray diffraction (XRD) analysis, and one can obtain large ($\sim$ 3 $\times$ 2 mm$^2$) platelet-shaped single crystals which were aligned by Laue analysis and  then cut along the $a$ or $b$ axis. 

The electrochemical intercalation of Cu into PSBS to obtain Cu$_{x}$(PbSe)$_{5}$(Bi$_{2}$Se$_{3}$)$_{6}$ (CPSBS) was done by following the technique reported in Ref. \cite{Kriener2011} in a saturated solution of CuI in acetonitrile CH$_3$CN. A bare Cu rod was used as the positive electrode and the sample suspended with a Cu wire as the negative electrode. The integrated charge while applying a current of $10~\mu$A was used for estimating the $x$ value, i.e., the total mol of the intercalated Cu. The $x$-value estimated this way was verfied by a direct measurement of the mass change using an ultra-high-resolution ($\pm$0.1 $\mu$g) balance, showing no systematic deviation from the charge-based measurement. To activate the superconductivity, the samples were annealed in evacuated SiO$_2$-glass tubes at 580\degree C for 3 hours and then quenched into ice water.

\begin{flushleft} 
{\bf S2. Experimental details}
\end{flushleft}

The resistivity measurements were performed using a four-probe ac lock-in technique with excitation currents of 0.5 -- 1 mA. The electrical contacts on the samples were made by using a vacuum-cure-type silver paint with the configuration shown in Fig. 1(a) of the main text. 
The specific-heat measurements were performed using a standard relaxation-time method. After stabilizing distinct sample temperatures, a constant current was applied through the platform heater until the sample temperature is raised by typically 75~mK, and then the current is switched off again. The recorded time-dependent temperature variation, $\Delta T(t)$, for both periods (heater on/off) was fitted by exponential functions and the total heat capacity was derived from the corresponding time constants $\tau$. The sample's specific heat was then calculated after subtracting the addenda's heat capacity which has been determined in a separate run measuring the empty platform. The home-built calorimeter, schematically shown in Fig. 1(a) of the main text, consists of the addenda made of a sapphire substrate, a CERNOX bare-chip thermometer, and an evaporated thin gold film showing 100~$\Omega$ at room temperature as a heater. The addenda is suspended by thin NbTi wires ($\sim$50 $\mu$m diameter) obtained by removing Cu clad from a superconducting magnet wire. The NbTi wires are glued to the addenda by using silver epoxy.

\begin{flushleft} 
{\bf S3. Subtraction of the phononic specific heat}
\end{flushleft}

\begin{figure}[b]
	\centering
	\includegraphics[width=0.45\linewidth]{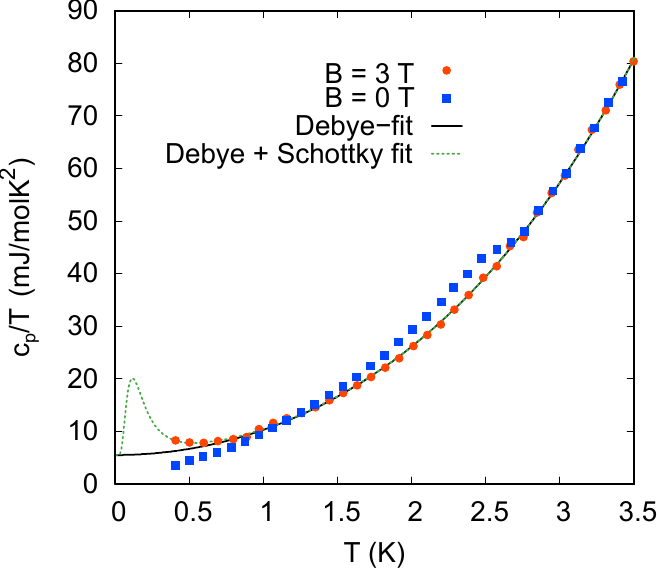}
	\caption{Specific heat $c_p(T)$ of sample B measured in 0 and 3 T. The solid line is a fit of the 3-T data to Eq. (1), and the dashed line additionally includes the Schottky contribution discussed in Sec. S3. The fitting parameters are listed in Table S1.}
\end{figure}

In order to separate the electronic specific heat $c_{\rm el}$ from the measured specific heat $c_p$, the phononic specific heat $c_{\rm ph}$ has to be subtracted. At sufficiently low temperature, $c_{\rm el}$ is linear in $T$, while $c_{\rm ph}$ shows mainly a $T^3$ behavior with a small $T^5$ contribution. The $c_p(T)$ data of our CPSBS samples in the normal state, which was induced by applying a magnetic field of 3 T along the $c^*$ axis, are well described by
\begin{equation}
c_p = c_{\rm el} + c_{\rm ph} =  \gamma T + \beta_1 T^3 + \beta_2 T^5,
\end{equation}
apart from a slight deviation at the lowest temperatures due to a Schottky anomaly.  
In Fig. S2 the $c_p(T)$ data of sample B measured in 0 and 3 T are shown, where the 3-T data are fitted by Eq. (1). Inclusion of the Schottky contribution (which is discussed in the next section) yields the dashed line. The electronic and phononic coefficients extracted from this fit are listed in Table S1 and compared to those reported in Ref. \cite{Sasaki2014}.

 \begin{table}[t]
	\begin{tabular}{c|c c c}
		Sample & \,\,$\gamma$~(mJ/molK$^2$) \,\,& $\beta_1$~(mJ/molK$^4$) \,\,& $\beta_2$~(mJ/molK$^6$) \\ 
		\hline 
		B & 5.80 & 4.12 &  0.20\\ 
		Sasaki {\it et al.} \cite{Sasaki2014}\, & 5.89  &3.73  &0.10  
\end{tabular} 
\caption{Fitting parameters of $c_p(T)$ in the normal state of CPSBS. }
\end{table}

\begin{flushleft} 
{\bf S4. Schottky anomaly}
\end{flushleft}

\begin{figure}[t]
	\centering
	\includegraphics[width=0.6\linewidth]{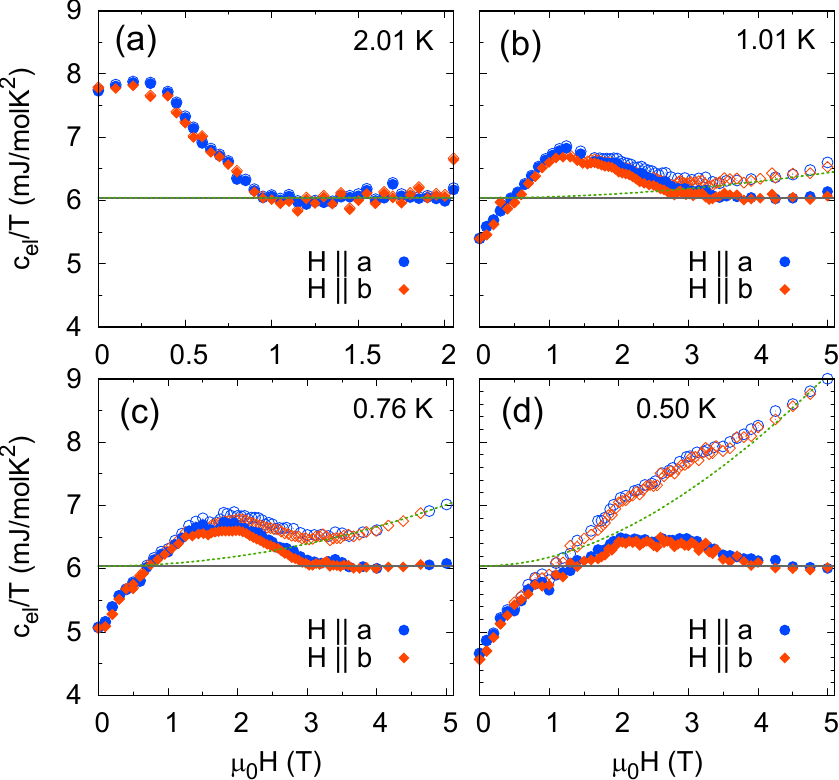}
	\caption{Magnetic-field dependencies of $c_{\rm el}/T$ measured at (a) 2.01 K, (b) 1.01 K, (c) 0.76 K, and (d) 0.50~K on sample B. The closed and open symbols, respectively, show the data with and without the Schottky correction calculated with $g$ = 0.17 and $n$ = 5.4~mJ/molK. The solid horizontal lines show the putative magnetic-field-independent normal-state values of $c_{\rm el}/T$ at respective temperatures; adding the calculated Schottky anomaly gives the dashed lines. Note that the Schottky anomaly is negligibly small a 2.01~K.}
\end{figure}

As was already reported in Ref.~\cite{Sasaki2014},  the $c_p(T)$ data of CPSBS show a  magnetic-field dependent Schottky anomaly with a very small $g$-factor of 0.17, whose origin is not fully clarified yet. To correctly isolate $c_{\rm el}$ coming from the conduction electrons, one needs to subtract this Schottky contribution, which systematically increases with the ratio $H/T$, i.e., with increasing field and decreasing temperature. The Schottky contribution to the specific heat, $c_{\rm Sch}$, is modelled by a two-level system corresponding to free $S = 1/2$ moments \cite{Emerson1994, Moler1997}
\begin{equation}
c_{\rm Sch}(T,B) = \frac{n x^2 e^x}{(1 + e^x)^2}
\quad \quad \left(x \equiv \frac{g \mu _B B}{k_B T}\right) \, .
\end{equation}
Here, the coefficient $n$ is proportional to the density of the magnetic impurities and $g$ is their gyromagnetic ratio. In the previous work on CPSBS with $x$ = 1.36, an excellent description of the observed anomaly was found for $n$ = 3.1~mJ/molK and $g$ = 0.17 \cite{Sasaki2014}. In the present work, we find the same $g$ = 0.17 and a slightly larger $n$ = 5.4~mJ/molK. Figure S3 shows the magnetic-field dependencies of $c_{\rm el}/T$ at 2.01, 1.01, 0.76, and 0.50~K before and after the Schottky correction calculated with these parameters. Also, the dashed line in Fig. S2 includes this Schottky correction, which well describes the temperature-dependent $c_p(T)$ data in 3 T.

\vspace{2cm}

\begin{flushleft} 
{\bf S5. Reproducibility of the two-fold in-plane symmetry in $H_{c2}$}
\end{flushleft} 

\begin{figure}[t]
	\centering
	\includegraphics[width=0.55\linewidth]{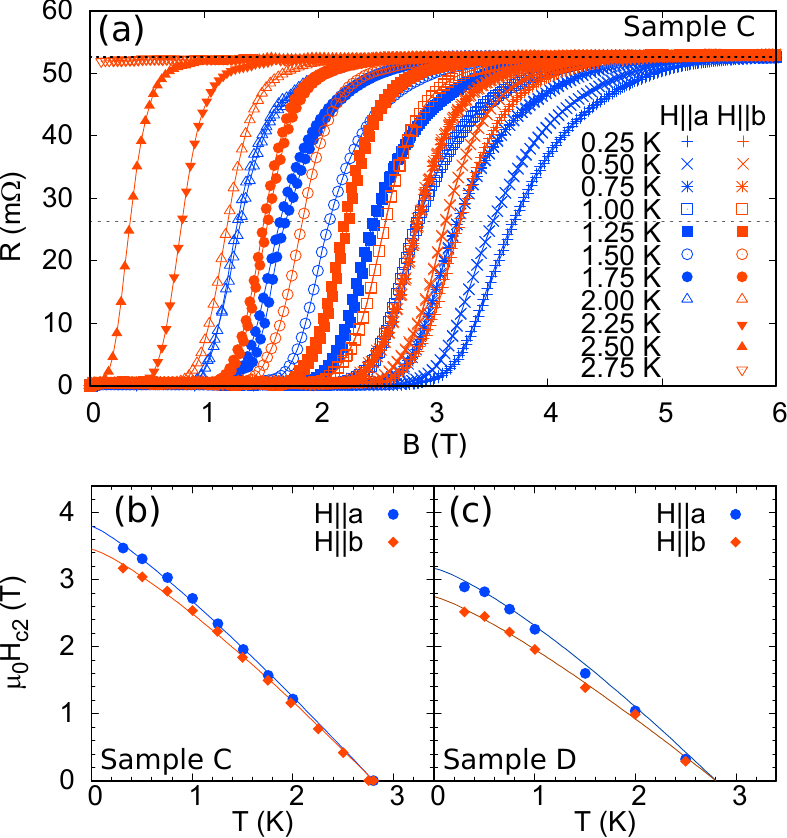}
	\caption{(a) $R_{\perp}(H)$ curves measured on sample C in $H \parallel a$ and $H \parallel b$ at various temperatures. (b) Temperature dependencies of $H_{c2}$ in sample C for $H \parallel a$ and $H \parallel b$ extracted from the data in panel (a) by taking the mid-point in the resistive transitions as the criterion. (c) Temperature dependencies of $H_{c2}$ in sample D for $H \parallel a$ and $H \parallel b$ extracted from $R_b(H)$ data.}
\label{fig:fig4}
\end{figure}

\begin{figure}[t]
	\centering
	\includegraphics[width=0.6\linewidth]{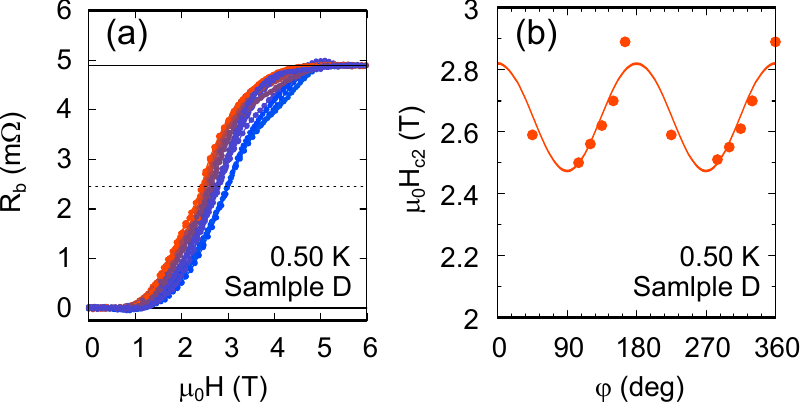}
	\caption{(a) $R_b(H)$ curves measured on sample D at 0.50 K for varying directions of the in-plane magnetic field. (b) The $\varphi$ dependence of $H_{c2}$ at 0.5 K extracted from the data in panel (a).  The angle $\varphi$ is measured from the $a$ axis.}
\end{figure}

\begin{figure}[t]
	\centering
	\includegraphics[width=0.6\linewidth]{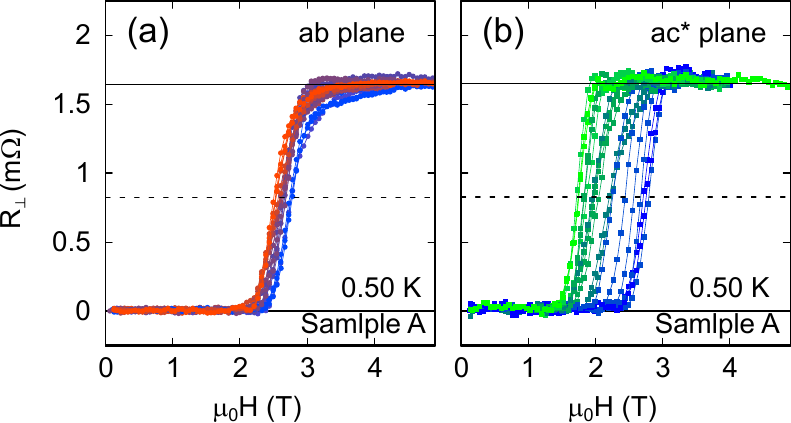}
	\caption{(a), (b) $R_{\perp}(H)$ curves measured on sample A at 0.50 K for varying directions of the magnetic field which was rotated in (a) the $ab$ plane and (b) the $ac^*$ plane. The depencencies $H_{c2}(\varphi)$ and $H_{c2}(\psi)$ shown in Fig. 2(b) of the main text were generated from these data.}
\end{figure}

In addition to samples A and B discussed in the main text, the in-plane anisotropy of $H_{c2}$ was measured in two more samples, C and D, by resistance measurements.
In sample C, the current was applied along the $c^*$ direction as in sample A, while the current was along the $b$ direction in sample D. The measurements of samples C and D were done in a solenoid magnet by using an {\it in-situ} piezo-rotator (Attoqube ANRv51) with a possible magnetic-field-direction misalignment of $\pm$2\degree.

Detailed $H_{c2}(T)$ behavior was extracted through a series of $R_{\perp}(H)$ measurements on sample C in $H \parallel a$ and $H \parallel b$ at various temperatures shown in Fig. S4(a). The resulting $H_{c2}(T)$ behavior for these two field orientations are shown in Fig. S4(b), which reproduces the finding shown in Fig. 4(c) of the main text. 

Also, a series of $R_b(H)$ curves with changing $\varphi$ were measured at 0.50 K [Fig. S5(a)] to extract the detailed dependence of $H_{c2}$ on the in-plane magnetic-field direction [Fig. S5(b)], which reproduces the two-fold symmetric change in $H_{c2}(\varphi)$ with maxima at $H \parallel a$. Hence, one can see that the two-fold in-plane symmetry is robust against the measurement current direction. The $H_{c2}(T)$ behavior of sample D for $H \parallel a$ and $H \parallel b$ are shown in Fig. S4(c).
One may notice that the $H_{c2}$ values obtained from the resistivity data by the mid-point  criteria on samples A and D well agree to each other, while larger values are obtained  for sample C. Such a sample dependence could result from different impurity contents, because a shorter residual mean free path typically reduces the superconducting coherence length and therefore increases the $H_{c2}$ value. Independent from this variation obtained for different samples, each individual data set consistently yields an in-plane $H_{c2}$-anisotropy of about 10\,\%, which is also obtained from the specific-heat data measured on sample B. 

For completeness, we show the raw data of the series of $R_{\perp}(H)$ curves measured on sample A at 0.50 K either with changing $\varphi$ (in-plane angle measured from the $a$ axis) or with changing $\psi$ (out-of-plane angle measured from the $a$ axis) in Figs. S6(a) and S6(b), respectively, from which the plots in Fig. 2(b) were generated. Table S2 gives an overview of all the measurements performed on the four samples used in this study. 

\vspace{5mm}
\begin{table}[h]
	\begin{tabular}{c|c c c c c c c}
		Sample	\,& \,\,\,$x$\, & \,$T_c$ (K)\, & \,SF (\%)\, & \, measurements\, & \,variables & $L_{\rm a}\times L_{\rm b}$ (mm$^2$)  & mass (mg)\\ 
		\hline 
		A &\,\,\, 1.47 & 2.87 & 75 & $R_{\perp}$ & $H$, $T$, $\varphi$, $\psi$ & 2.2 $\times$ 8.2 & 17.6210 \\ 
		B &\,\,\, 1.47 & 2.82 & 104 & $c_p$ & $H$, $T$, $\varphi$ & 2.6 $\times$ 3.8 & 17.3250 \\ 
		C &\,\,\, 1.34 & 2.85 & 89 & $R_{\perp}$ & $H$, $T$, $\varphi$ & 2.6 $\times$ 2.9 & 7.8783\\ 
		D &\,\,\, 1.49 & 2.82 & 113 & $R_b$ & $H$, $T$, $\varphi$ & 2.1 $\times$ 3.2 & 15.4659
	\end{tabular}
	\caption{Overview of the samples used in this work. ``SF'' is the shielding fraction at 1.8 K. $L_a$ and $L_b$ are the lengths of the edges of the samples in the $a$ and $b$ axes, respectively.}
\end{table}

\begin{flushleft} 
{\bf S6. Comparison of the crystal structures of Bi$_2$Se$_3$ and (PbSe)$_5$(Bi$_2$Se$_3$)$_6$}
\end{flushleft} 

In Fig.~S7, we clarify the geometrical relation between the crystal structures of Bi$_2$Se$_3$ and (PbSe)$_5$(Bi$_2$Se$_3$)$_6$. The structural information is based on the crystallographic data measured and refined by Vicente {\it et al.} for Bi$_2$Se$_3$ \cite{Vicente1999} and by Fang {\it et al.} for PSBS  \cite{Fang2014}. The principal structural units, which are present in both systems, are the Bi$_2$Se$_3$ quintuple-layers (QLs) consisting of individual Se--Bi--Se--Bi--Se layers. These QLs are stacked along the crystallographic $c$ axis in pristine Bi$_2$Se$_3$, which has rhombohedral symmetry and crystallizes in the space group $R\bar{3}m$. Using the hexagonal notation to define the unit cell, the $a$ and $b$ axes are crystallographically equivalent and lie within the QL plane with a length equal to the  distance of nearest-neighbor Se (or Bi) atoms within the individual planes. The $c$ axis is perpendicular to these planes and has a length of three QLs. The $\bar{3}$ rotation axis is parallel to $c$ generating three equivalent mirror planes, that are perpendicular to the $a$ ($b$) axis, and three equivalent two-fold rotation axes, that are perpendicular to the mirror planes. 

\begin{figure}[t]
	\centering
	\includegraphics[width=0.9\linewidth]{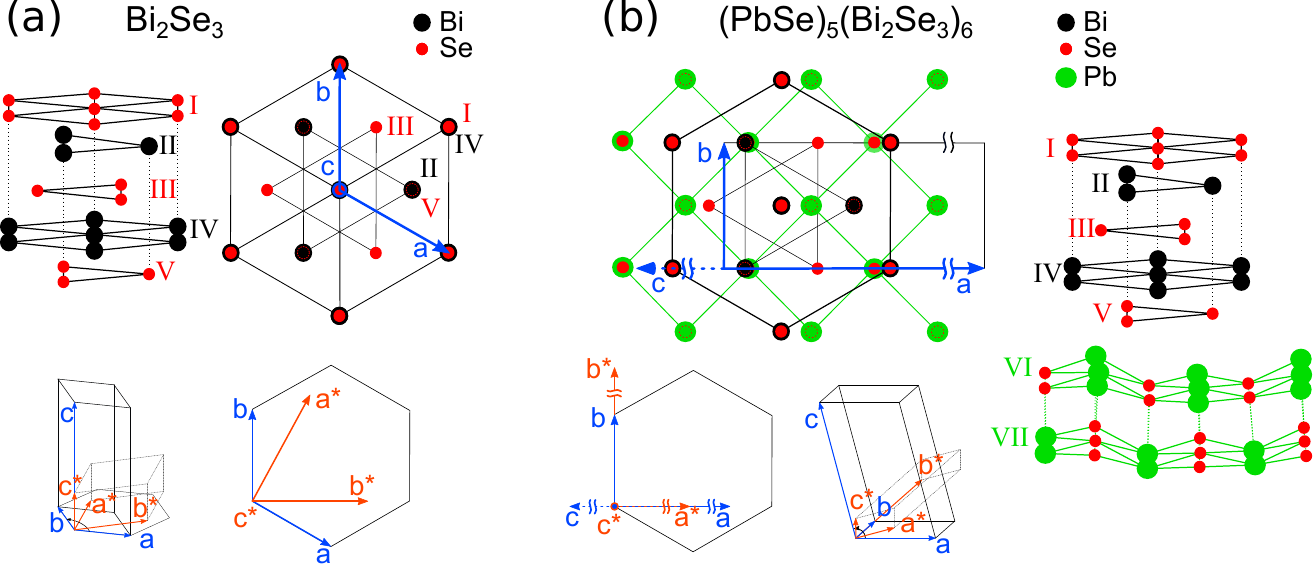}
	\caption{Comparison of the crystal structures of (a) Bi$_2$Se$_3$ and (b) (PbSe)$_5$(Bi$_2$Se$_3$)$_6$. The upper row sketches the stackings of the various layers along the $c$ axis and shows the orientations of the in-plane $a$ and $b$ axes. The lower row displays the crystallographic unit cells in real and in reciprocal space for both structures and also displays the projections of $\bf{a}$, $\bf{b},$ $\bf{c}$ and of $\bf{a}^*$, $\bf{b}^*$, $\bf{c}^*$ onto the $ab$ planes. Note that the hexagonal $a$ axis and the monoclinic $a$ axes have different orientations.}
	\label{fig:fig7}
\end{figure}

In (PbSe)$_5$(Bi$_2$Se$_3$)$_6$, the stacking along $c$ is such that two Bi$_2$Se$_3$ QLs alternate with one PbSe bilayer (BL). Due to a mismatch between the Bi$_2$Se$_3$ QLs and the PbSe BLs, the symmetry of (PbSe)$_5$(Bi$_2$Se$_3$)$_6$ is reduced to $C2/m$. The close relationship between the two materials is reflected in the fact that the $C2/m$ space group is a subgroup of $R\bar{3}m$, so that all symmetry elements of (PbSe)$_5$(Bi$_2$Se$_3$)$_6$ are already present in Bi$_2$Se$_3$. The $\bar{3}$ rotation axis is, however, destroyed by the presence of the PbSe BLs and only one mirror plane and one perpendicular two-fold rotation axis survive.   

For (PbSe)$_5$(Bi$_2$Se$_3$)$_6$, it is convenient to choose the unique monoclinic $b$ axis to have the same direction and length as the hexagonal $b$ axis of Bi$_2$Se$_3$. For an easy comparison of the two structures, the monoclinic $a$ axis can also be chosen within the QL plane, leaving the $c$ axis pointing out of the plane with the monoclinic angle $\beta$ = 107.224$^{\circ}$ to the $a$ axis. Note that the monoclinic $b$ axis is the 2-fold rotation axis and perpendicular to the mirror plane of (PbSe)$_5$(Bi$_2$Se$_3$)$_6$. In simple words, the symmetry reduction from rhombohedral to monoclinic specifies one out of the three originally equivalent in-plane directions of the Bi$_2$Se$_3$ QLs. Consequently, the monoclinic $b$ axis of (PbSe)$_5$(Bi$_2$Se$_3$)$_6$ is parallel to the original hexagonal $b$ axis of Bi$_2$Se$_3$, whereas the monoclinic $a$ axis is perpendicular to this direction.  

In order to prevent possible confusions, it is worth mentioning that in the standard cell for the space group $C2/m$, the $a$ and $c$ axes are switched \cite{inttablesforcrystallographyVolA}. As a consequence, the centering position for the definition used here is found on the $A$ instead of the $C$ face. This becomes important, for example, when considering the reflection conditions reported for the (00$l$) plane by Fang {\it et al.} \cite{Fang2014}. We also note that Fig.~S7 shows the reciprocal lattice vector  ${\bf c}^* || ({\bf a} \times {\bf b})/V_{uc}$, whereas some previous reports~\cite{Segawa2015,Sasaki2014} defined a real-space $\bf{c}^*$ with the same orientation and a length $|{\bf c}^*| = |{\bf c}| \cos(\beta)$.   

Finally, we would like to emphasize that despite the lower symmetry of (PbSe)$_5$(Bi$_2$Se$_3$)$_6$ compared to Bi$_2$Se$_3$, the structural distortion of the Bi$_2$Se$_3$ QL units is actually weak. This is shown in Fig.~S8, which summarizes the bond lengths and interior angles of a Se hexagon of (PbSe)$_5$(Bi$_2$Se$_3$)$_6$.

\begin{figure}[t]
	\centering
	\includegraphics[width=0.4\linewidth]{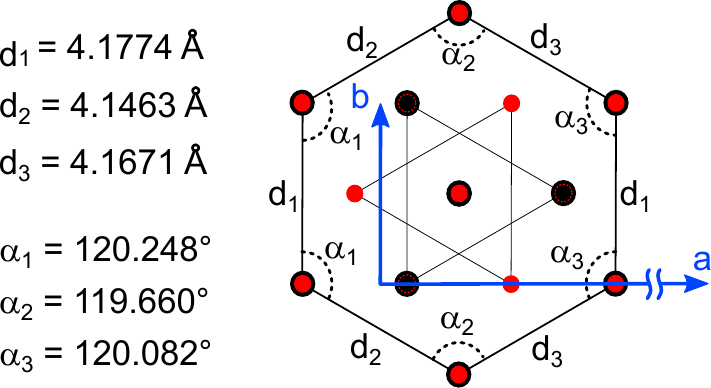}
	\caption{Lengths and  interior angles of a Se hexagon in (PbSe)$_5$(Bi$_2$Se$_3$)$_6$.}
\end{figure}


\end{document}